\documentstyle[psfig]{aipproc}
\def\nn{\noindent}
\def\ie{{\it i.e.}}
\def\eg{{\it e.g.}}
\def\etc{{\it etc}}
\def\etal{{\it et al.}}

\def\epem{\ifmmode e^+e^-\else $e^+e^-$\fi}
\def\to{\rightarrow}

\def\mpl{\ifmmode \overline M_{Pl}\else $\bar M_{Pl}$\fi}

\begin{document}

\rightline{\vbox{\halign{&#\hfil\cr
SLAC-PUB-8297\cr
November 1999\cr}}}
\vspace{0.8in}

\title{{Kaluza-Klein Physics at\\ Muon Colliders}
\footnote{To appear in the {\it Proceedings of the Study on Colliders and 
Collider Physics at the Highest Energies: Muon Colliders at 10 TeV to 
100 TeV}, Montauk Yacht Club Resort, Montauk, New York, 
27 September--1 October 1999}
}

\author{Thomas G Rizzo
\footnote{
E-mail:rizzo@slacvx.slac.stanford.edu. 
Work supported by the Department of Energy, 
Contract DE-AC03-76SF00515}
}

\address{Stanford Linear Accelerator Center\\
Stanford CA 94309, USA}

\maketitle

\begin{abstract}
We discuss the physics of Kaluza-Klein excitations of the Standard Model 
gauge bosons that can be explored by a high energy muon collider in the era 
after the LHC and TeV Linear Collider. We demonstrate that the muon collider 
is a necessary ingredient in the unraveling the properties of such states 
and, perhaps, proving their existence. The possibility of observing the 
resonances associated with the excited KK graviton states of the 
Randall-Sundrum model is also discussed. 
\end{abstract}

\section*{Introduction}

In theories with extra dimensions, $d\geq 1$, the gauge fields of the 
Standard Model(SM) will have Kaluza-Klein(KK) excitations if they are allowed 
to propagate in the bulk of the extra dimensions. If such a scenario is 
realized then, level by level, the masses of the excited states of the photon, 
$Z$, $W$ and gluon would form highly degenerate towers. The possibility that 
the masses of the lowest lying of these states, of order the inverse size of 
the compactification radius $\sim 1/R$, could be as low as 
$\sim$ a few TeV or less leads 
to a very rich and exciting phenomenology at future and, possibly, existing 
colliders{\cite {old}}. For the case of one extra dimension compactified on 
$S^1/Z_2$ the spectrum of the excited states is given by $M_n=n/R$ and the 
couplings of the excited modes relative to the corresponding zero mode to 
states remaining on the wall at the orbifold fixed points, such as the SM 
fermions, is simply $\sqrt 2$ for all $n$. These masses and couplings are 
insensitive to the choice of compactification in the case of one extra 
dimension assuming the metric tensor factorizes, \ie, the elements of the 
metric tensor on the wall are independent of the compactified co-ordinates. 

If such KK states exist what is the lower bound on their mass? We already know 
from direct $Z'/W'$ and dijet bump searches at the Tevatron from Run I that 
they must lie above $\simeq 0.85$ TeV{\cite {tev}}. A null 
result for a search made with 
data from  Run II will push this limit to $\simeq 1.1$ TeV or so. To do 
better than this at present we must rely on the indirect effects associated 
with KK tower exchange in what essentially involves a set of dimension-six 
contact interactions. Such limits rely 
upon a number of additional assumptions, in particular, that the effect of KK 
exchanges is the {\it only} new physics beyond the SM.  The strongest and least 
model-dependent of these bounds arises from an analysis of charged current 
contact interactions at both HERA and the Tevatron by Cornet, Relano and 
Rico{\cite {cornet}} who, in the case of one extra dimension, obtain a bound 
of $R^{-1}>3.4$ TeV. Similar analyses have been carried out by a number of 
authors{\cite {host,rw}}; the best limit arises from an updated combined fit 
to the precision electroweak data{\cite {rw}} as presented at the 1999 summer 
conferences{\cite {data}} and yields{\cite {kktest}} $R^{-1}>3.9$ TeV for the 
case of one extra dimension. 
From the previous discussion we can also draw a further conclusion for the 
case $d=1$: the lower bound $M_1>3.9$ TeV is so strong that the {\it second} KK 
excitations, whose masses must now exceed 7.8 TeV due to the above scaling 
law, will be beyond the reach of 
the LHC. This leads to the important result that the LHC will {\it at most} 
only detect the first set of KK excitations for $d=1$.

In all analyses that obtain indirect limits on $M_1$, one is actually 
constraining a dimensionless quantity such as 
\begin{equation}
V=\sum_{{\bf n}=1}^\infty {g_{\bf n}^2\over {g_0^2}} 
{M_w^2\over {M_{\bf n}^2}} \,,
\end{equation}
where, generalizing the case to $d$ additional dimensions, $g_{\bf n}$ is the 
coupling and $M_{\bf n}$ the mass of the 
$n^{th}$ KK level labelled by the set of $d$ integers {\bf n} and $M_w$ 
is the $W$ boson mass which we employ as a typical weak scale factor. For 
$d=1$ this sum is finite since $M_n=n/R$ and $g_n/g_0=\sqrt 2$ for $n>1$; 
one immediately 
obtains $V={\pi^2\over {3}}(M_w/M_1)^2$ with $M_1$ being the mass of the 
first KK excitation. From the precision data one obtains a bound on $V$ and 
then uses the above expression to obtain the corresponding bound on $M_1$. 
For $d>1$, however, independently of how the extra 
dimensions are compactified, the above sum in $V$ {\it diverges} and so it 
is not so straightforward to obtain a bound on $M_1$. We also 
recall that for $d>1$ the mass spectrum and the relative coupling strength of 
any particular KK excitation now become dependent upon how the additional 
dimensions are compactified. 

There are 
several ways one can deal with this divergence: ($i$) The simplest approach 
is to argue that as the states being summed in $V$ get heavier they approach 
the mass of the string scale, $M_s$, above which we know little and some new 
theory presumably takes over. Thus we 
should just truncate the sum at some fixed maximum value $n_{max}\simeq M_sR$ 
so that masses KK masses above $M_s$ do not contribute. 
($ii$) A second possibility is to 
note that the wall on which the SM fermions reside is not completely rigid 
having a finite tension. The authors in Ref.{\cite {wow}} argue that this wall 
tension can act like an exponential suppression of the couplings of the 
higher KK states in the tower thus rendering the summation finite, \ie, 
$g_{\bf n}^2 \to g_{\bf n}^2 e^{-(M_n/M_1)^2/n_{max}^2}$, where 
$n_{max}$ now parameterizes the strength of the exponential cut-off. 
(Antoniadis{\cite {kktest}} has argued that such an exponential suppression 
can also arise from considerations of string scattering amplitudes at 
high energies.) For a fixed value of $n_{max}$, the exponential approach is 
found to be more effective and lead to a smaller sum than that obtained by 
simple truncation and thus to a weaker bound on $M_1$. ($iii$) A 
last scenario{\cite {schm}} is to note the possibility that the SM wall 
fermions may have a finite size in the extra dimensions which smear out and 
soften the couplings appearing in the sum to yield a finite result. In this 
case the suppression is also of the Gaussian variety. 

We note 
that in all of the above approaches the value of the sum increases rapidly 
with $d$ for a fixed value of the cut-off parameter $n_{max}$. For $d=2(>2)$ 
the sum behaves asymptotically as $\sim log ~n_{max} (\sim n_{max}^{d-2})$. 
This leads to the very important result that, for a fixed bound on $V$ from 
experimental data, the corresponding bound on the mass of the lowest lying KK 
excitation rapidly 
strengthens with the number of extra dimension, $d$. Table I shows 
how the $d=1$ lower bound of 3.9 TeV for the mass of $M_1$ changes as we 
consider different compactifications for $d>1$. We see that in some cases the 
value of $M_1$ is so large it will be beyond the mass range accessible to the 
LHC as it is for all cases of the $d=3$ example.

\vspace*{0.3cm}
\begin{table*}[htpb]
\caption{Lower bound on the mass of the first KK state in TeV resulting from 
the constraint on $V$ for the case of more than one dimension. `T'[`E'] labels 
the result 
obtained from the direct truncation (exponential suppression). Cases 
labeled by an asterisk will be observable at the LHC. $Z_2\times Z_2$ and 
$Z_{3,6}$ correspond to compactifications in the case of $d=2$ 
while $Z_2\times Z_2\times Z_2$ is for the case of $d=3$.}
\begin{tabular}{lcccccc}
     &\multicolumn{2}{c}{ $Z_2\times Z_2$}  &\multicolumn{2}{c}{$Z_{3,6}$} & 
\multicolumn{2}{c}{$Z_2\times Z_2\times Z_2$} \\ 
\tableline
\tableline
$n_{max}$ &  T   &  E   &  T   &  E   &  T   &  E    \\
\tableline
2   & 5.69$^*$  & 4.23$^*$  &6.63$^*$  &4.77$^*$ &8.65  & 8.01  \\  
3   & 6.64   & 4.87$^*$ & 7.41 &5.43$^*$ & 11.7 & 10.8 \\
4   & 7.20   & 5.28$^*$ & 7.95 &5.85$^*$ & 13.7 & 13.0 \\
5   & 7.69   & 5.58$^*$ & 8.36 &6.17$^*$ & 15.7 & 14.9 \\
10  & 8.89   & 6.42     & 9.61 &7.05  & 23.2  &  22.0  \\
20  & 9.95   & 7.16     & 10.2 &7.83  & 33.5  &  31.8  \\
50  & 11.2   & 8.04     & 12.1 &8.75  & 53.5  &  50.9  \\
\end{tabular}
\end{table*}
\vspace*{0.4cm}

\section*{SM KK States at the LHC and Linear Colliders}

Let us return to the $d=1$ case at the LHC where the degenerate KK states 
$\gamma^{(1)}$, $Z^{(1)}$, $W^{(1)}$ and $g^{(1)}$ are potentially visible. 
It has been shown {\cite {kktest}} that for masses in excess of $\simeq 4$ 
TeV the $g^{(1)}$ resonance in dijets will be washed out due to its rather 
large width and the experimental jet energy resolution available at the LHC 
detectors. Furthermore, 
$\gamma^{(1)}$ and $Z^{(1)}$ will appear as a {\it single} resonance in 
Drell-Yan that cannot be resolved and 
looking very much like a single $Z'$. Thus 
if we are lucky the LHC will observe what appears to be a degenerate $Z'/W'$. 
How can we identify these states as KK excitations when we remember that the 
rest of the members of the tower are too massive to be produced? We remind 
the reader that 
many extended electroweak models{\cite {models}} exist which predict a 
degenerate $Z'/W'$. Without further information, it would seem likely that 
this would become the most likely guess of what had been found.

%
\nn
\begin{figure}[htbp]
\centerline{
\psfig{figure=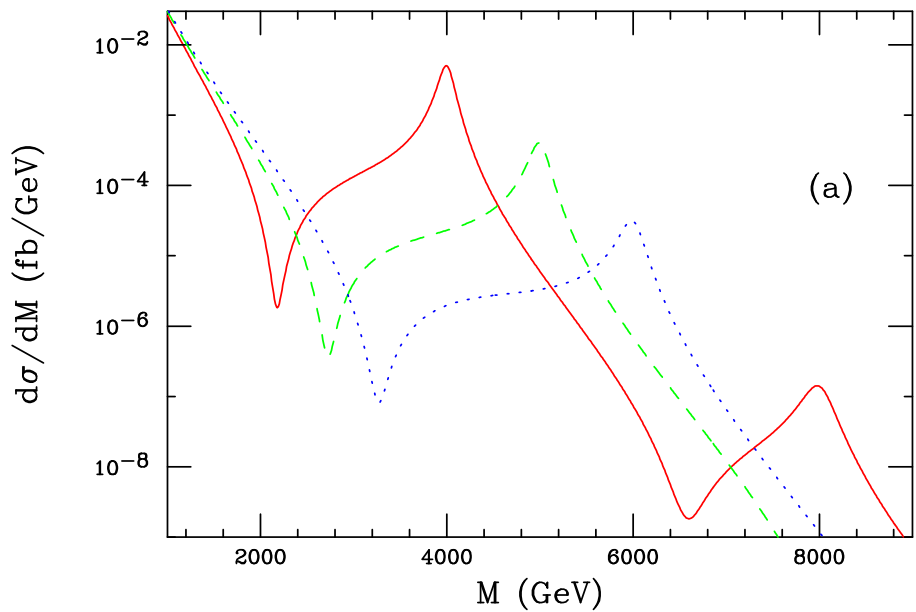,height=8.0cm,width=12cm,angle=90}}
\vspace*{0.1cm}
\caption{Cross section for Drell-Yan 
production of the degenerate neutral KK excitations $Z^{(n)}$ and 
$\gamma^{(n)}$ as a function of the dilepton invariant mass at the LHC 
assuming one extra dimension and naive coupling values with $1/R$=4(5, 6) TeV 
corresponding to the solid(dashed, dotted) curve. The second excitation is 
only shown for the case of $1/R=4$ TeV.}
\end{figure}
\vspace*{0.4mm}

To clarify this situation let us consider the results displayed in 
Figs. 1 for $d=1$ where we show the 
production cross sections in the $\ell^+\ell^-$ channel with inverse 
compactification radii of 4, 5 and 6 TeV. In calculating these cross sections 
we have assumed that the KK excitations have their naive couplings and 
can only decay to the usual fermions of 
the SM. Additional decay modes can lead to appreciably lower cross sections so 
that we cannot use the peak heights to determine the degeneracy of the KK 
state. Note that in the 4 TeV case, which is essentially as small a mass 
as can be tolerated 
by the present data on precision measurements, the second KK 
excitation is visible in the plot. We see several things from these figures. 
First, we can easily estimate the total number of events in 
the resonance regions associated with each of the peaks assuming the canonical 
integrated luminosity of $100 fb^{-1}$ appropriate for the LHC; we find 
$\simeq 300(32, 3, 0.02)$ events corresponding to the 4(5,6,8) TeV 
resonances if we sum over both electron and muon final states and assume 
$100\%$ leptonic 
identification efficiencies. Clearly the 6 and 8 TeV resonances will 
not be visible at the LHC (though a modest increase in luminosity by a factor 
of a few will allow 
the 6 TeV resonance to become visible) and we also verify our claim that 
only the first KK 
excitations will be observable. In the case of the 4 TeV resonance there is 
sufficient statistics that the KK mass will be well measured and 
one can also imagine measuring the forward-backward asymmetry, 
$A_{FB}$, if not the full 
angular distribution of the outgoing leptons, since the 
final state muon charges can be signed. Given sufficient statistics, a 
measurement of the angular distribution would demonstrate that the state 
is indeed spin-1 and not spin-0 or spin-2. However, for such a heavy resonance 
it is unlikely that much further information could be obtained about its 
couplings and other properties. In fact the conclusion of several years of $Z'$ 
analyses{\cite {snow}} 
is that coupling information will be essentially impossible to obtain for 
$Z'$-like resonances with masses in excess of 1-2 TeV at the LHC due to low 
statistics. 
Furthermore, the lineshape of the 4 TeV resonance and the Drell-Yan spectrum 
anywhere close to the peak will be difficult to measure 
in detail due to both the limited statistics and energy smearing. Thus we 
will never know from LHC data alone whether the first KK resonance has been 
discovered or, instead, some extended gauge model scenario has been realized. 
To make further progress we need a lepton collider.

It is well-known that future $e^+e^-$ linear colliders(LC) operating in the 
center of mass energy range $\sqrt s=0.5-1.5$ TeV will be sensitive to indirect 
effects arising from the exchange of new $Z'$ bosons with masses typically 6-7 
times greater than $\sqrt s${\cite {snow}}. This sensitivity is even greater 
in the case of KK excitations since towers of both $\gamma$ and $Z$ exist 
all of which have couplings larger than their SM zero modes. Furthermore, 
analyses have shown 
that with enough statistics the couplings of the new $Z'$ to the SM fermions 
can be extracted{\cite {coupl}} in a rather precise manner, especially when 
the $Z'$ mass is already approximately known from elsewhere, \eg, the LHC. (If 
the $Z'$ mass is not known 
then measurements at several distinct values of $\sqrt s$ can be used to 
extract both the mass as well as the corresponding couplings{\cite {me}}.) 
In the present situation, we imagine that the LHC has 
discovered and determined the mass of a $Z'$-like resonance in the 4-6 TeV 
range. Can the LC tell us anything about this object?

%
\nn
\begin{figure}[htbp]
\centerline{
\psfig{figure=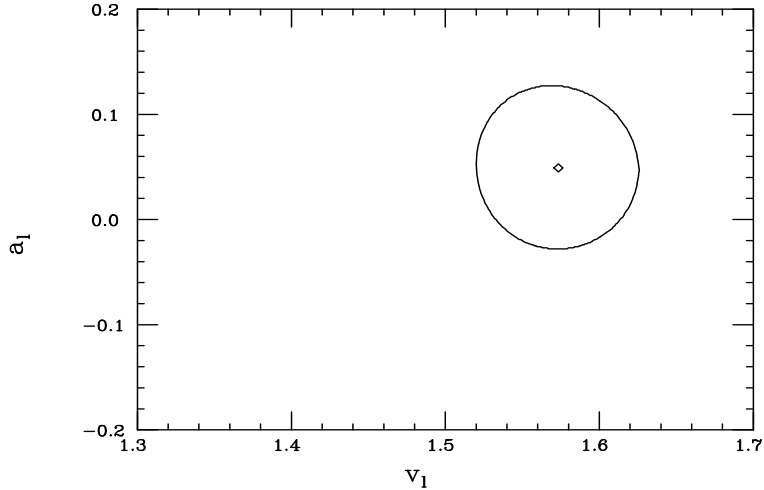,height=8cm,width=12cm,angle=-90}}
\vspace*{0.1cm}
\caption[*]{Fitted values of the parameters $v_l$ and $a_l$ following the 
procedures described in the text for a 4 TeV KK excitation at a 500 GeV  
$e^+e^-$ collider. The contour described the $95\%$ CL region with the best 
fit value as a diamond. The normalization is such that the corresponding SM 
$Z$ boson's axial-vector coupling to the electron is -1/2.}
\end{figure}
\vspace*{0.4mm}

The obvious step would be to use the LC to extract the couplings of the 
apparent resonance discovered by the LHC; we find that it is sufficient for 
our arguments below to do this solely for the leptonic channels. The idea is 
the following: we measure the deviations in the differential cross sections 
and angular dependent Left-Right polarization asymmetry, $A_{LR}^\ell$,
for the three lepton generations and combine those with $\tau$ polarization 
data. Assuming lepton universality(which would be observed in the LHC data 
anyway), that the resonance mass is well 
determined, and that the resonance is an ordinary $Z'$ we perform a fit to 
the hypothetical $Z'$ coupling to leptons, $v_l,a_l$. To be specific, let us 
consider the case of only one extra dimension with a 
4 TeV KK excitation and employ a $\sqrt s=500$ GeV 
collider with an integrated luminosity of 200 $fb^{-1}$. The result of 
performing this fit, including the effects of cuts and initial state 
radiation, is shown 
in Fig.2. Here we see that the coupling values are `well determined'
(\ie, the size of the $95\%$ CL 
allowed region we find is quite small) by the fitting 
procedure as we would have expected from previous analyses of $Z'$ couplings
extractions at linear colliders{\cite {snow,coupl,me}}.

%
\nn
\begin{figure}[htbp]
\centerline{
\psfig{figure=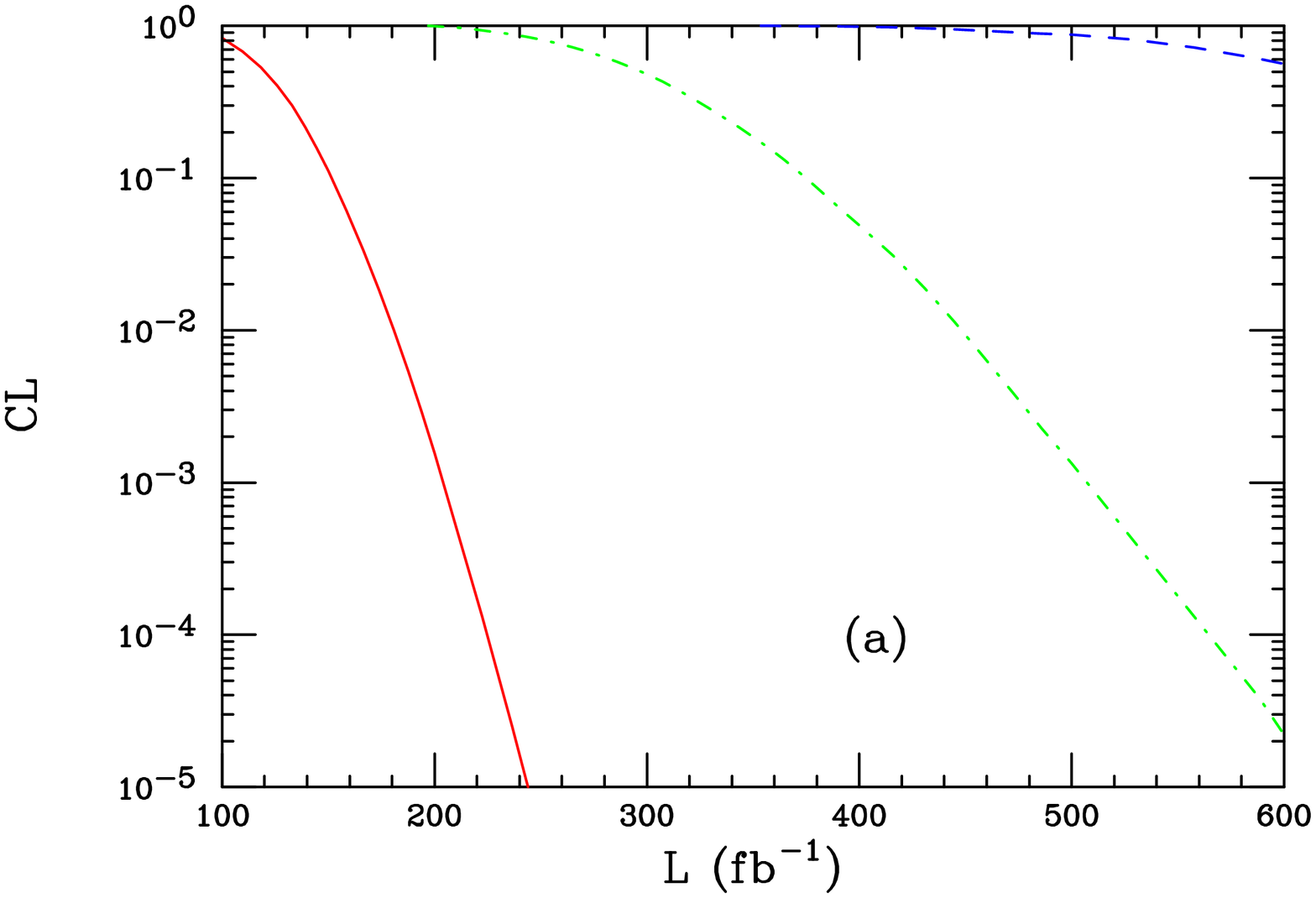,height=8.0cm,width=12cm,angle=90}}
\vspace*{9mm}
\centerline{
\psfig{figure=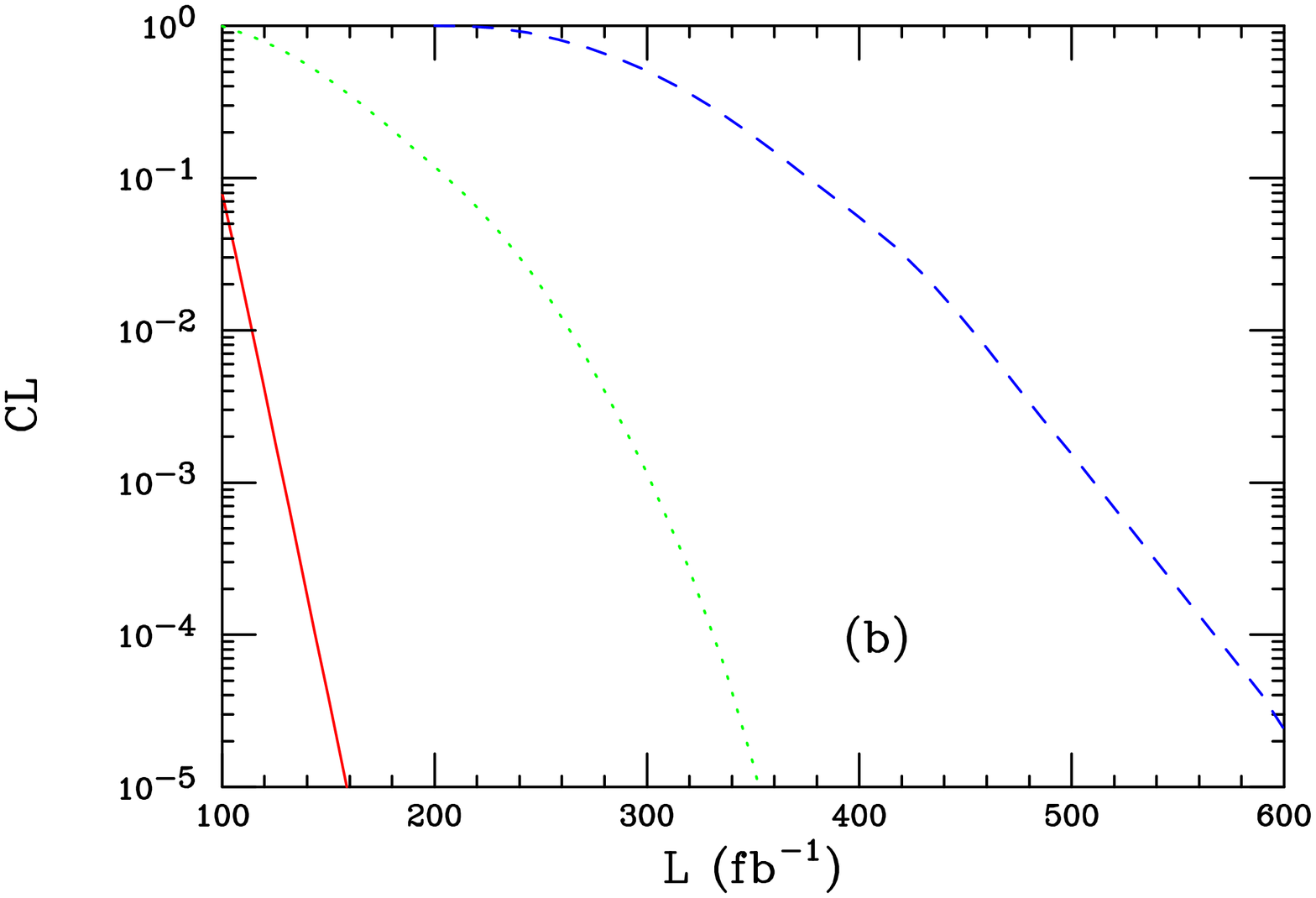,height=8.0cm,width=12cm,angle=90}}
\vspace*{0.0cm}
\caption{CL as a function of the integrated luminosity 
resulting from the coupling fits following from the analysis 
discussed in the text for both (a) a 500 GeV or a (b) 1 TeV $e^+e^-$ collider. 
In (a) the solid(dash-dotted,dotted) curve corresponds to a first KK 
excitation mass of 4(5,6) TeV. In (b) the solid(dotted,dashed) curve 
corresponds to a first KK mass of 5(6,7) TeV.}
\end{figure}
\vspace*{0.4mm}

The only problem with the fit shown in the figure is that 
the $\chi^2$ is very large leading to a very small confidence level, \ie, 
$\chi^2/d.o.f=95.06/58$ or CL=$1.55\times 10^{-3}$! (We note that this result 
is not very sensitive to the assumption of $90\%$ beam polarization; $70\%$ 
polarization leads to almost identical results.) For an ordinary $Z'$ it has 
been shown that fits of much higher quality, based on confidence level values, 
are obtained by this same procedure. 
Increasing the integrated luminosity can be seen to 
only make matters worse. Fig.3 shows 
the results for the CL following the above approach as we vary both the 
luminosity and the mass of the first KK excitation at both 500 GeV and 1 TeV 
$e^+e^-$ linear colliders. From this figure we see that the resulting CL 
is below $\simeq 10^{-3}$ for a first KK excitation with a mass of 4(5,6) 
TeV when the 
integrated luminosity at the 500 GeV collider is 200(500,900)$fb^{-1}$ whereas 
at a 1 TeV  for excitation masses of 5(6,7) TeV we require luminosities of 
150(300,500)$fb^{-1}$ to realize this same CL. Barring some unknown systematic 
effect the only conclusion that one could draw from such bad fits is that the 
hypothesis of a single $Z'$, and the existence of no other new physics, 
is simply {\it wrong}.  
If no other exotic states are observed below the first KK mass at the LHC, 
such as $\tilde \nu${\cite {rp}} or leptoquarks{\cite {leptos}}, this 
result would give very strong indirect evidence that something more unusual 
that a conventional $Z'$ had been found but {\it cannot} prove that this is a 
KK state.

\section*{SM KK States at Muon Colliders}

In order to be completely sure of the nature of the first KK excitation, we 
must produce it directly at a higher energy lepton collider and sit on and 
near the peak of the KK resonance. To reach this mass range will most likely 
require a Muon Collider.
The first issue to address is the quality of the degeneracy 
of the $\gamma^{(1)}$ and $Z^{(1)}$ states. 
Based on the analyses in Ref.{\cite {host,rw}} we can get 
an idea of the maximum possible size of this fractional mass shift and we 
find it to be 
of order $\sim M_Z^4/M_{Z^{(1)}}^4$, an infinitesimal quantity for KK masses 
in the several TeV range. Thus even when mixing is included we find that the 
$\gamma^{(1)}$ and $Z^{(1)}$ states remain very highly degenerate so that even 
detailed lineshape measurements may not be able to 
distinguish the $\gamma^{(1)}/Z^{(1)}$ composite state from that of a $Z'$. 
We thus must turn to other parameters in order to separate these two cases.

Sitting on the resonance there are a very large number of quantities that can 
be measured: the mass and apparent 
total width, the peak cross section, various partial 
widths and asymmetries \etc. From the $Z$-pole studies at SLC and LEP, we 
recall a few important tree-level results which we would expect to apply 
here as well 
provided our resonance is a simple $Z'$. First, we know that the value of 
$A_{LR}=[A_e=2v_ea_e/(v_e^2+a_e^2)]$, as measured on the $Z$ by SLD, does not 
depend on the fermion flavor of the final state and second, that the 
relationship $A_{LR}\cdot A_{FB}^{pol}(f)=A_{FB}^f$ holds, where 
$A_{FB}^{pol}(f)$ is the polarized Forward-Backward asymmetry as measured for 
the $Z$ at SLC and $A_{FB}^f$ is the usual Forward-Backward asymmetry. The 
above relation is seen to be trivially satisfied on the $Z$(or on a $Z'$) since 
$A_{FB}^{pol}(f)={3\over 4}A_f$ and $A_{FB}^f={3\over 4}A_eA_f$. Both of these 
relations are easily shown to fail in the present case of a `dual' resonance 
though they will hold if only one particle is resonating.

A short exercise 
shows that in terms of the couplings to $\gamma^{(1)}$, which we will call 
$v_1,a_1$, and $Z^{(1)}$, now called $v_2,a_2$, these same observables can 
be written as 
\begin{eqnarray}
A_{FB}^f &=& {3\over 4} {A_1\over D}\nonumber \\
A_{FB}^{pol}(f) &=& {3\over 4} {A_2\over D}\nonumber \\
A_{LR}^f &=& {A_3\over D}\,,
\end{eqnarray}
where $f$ labels the final state fermion and we have defined the coupling 
combinations
\begin{eqnarray}
D &=& (v_1^2+a_1^2)_e(v_1^2+a_1^2)_f+R^2(v_2^2+a_2^2)_e(v_2^2+a_2^2)_f
\nonumber\\
& & \quad +2R(v_1v_2+a_1a_2)_e(v_1v_2+a_1a_2)_f\\
A_1 &=& (2v_1a_1)_e(2v_1a_1)_f+R^2(2v_2a_2)_e(2v_2a_2)_f+2R(v_1a_2+v_2a_1)_e
(v_1a_2+v_2a_1)_e\nonumber \\
A_2 &=& (2v_1a_1)_f(v_1^2+a_1^2)_e+R^2(2v_2a_2)_f(v_2^2+a_2^2)_e+2R(v_1a_2
+v_2a_1)_f(v_1v_2+a_1a_2)_e\nonumber \\
A_3 &=& (2v_1a_1)_e(v_1^2+a_1^2)_f+R^2(2v_2a_2)_e(v_2^2+a_2^2)_f+2R(v_1a_2
+v_2a_1)_e(v_1v_2+a_1a_2)_f\,,\nonumber
\end{eqnarray}
with $R$ being the ratio of the widths of the two KK states, 
$R=\Gamma_1/\Gamma_2$, and the $v_{1,2i},a_{1,2i}$ are 
the appropriate couplings for electrons and fermions $f$. Note that when $R$ 
gets either very large or very small we recover the usual `single resonance' 
results. 
Examining these equations we immediately note that $A_{LR}^f$ is 
now {\it flavor dependent} and that the relationship between observables is 
no longer satisfied:
\begin{equation}
A_{LR}^f\cdot A_{FB}^{pol}(f)\neq A_{FB}^f\,,
\end{equation}
which clearly tells us that we are actually producing more than one resonance.

%
\nn
\begin{figure}[htbp]
\centerline{
\psfig{figure=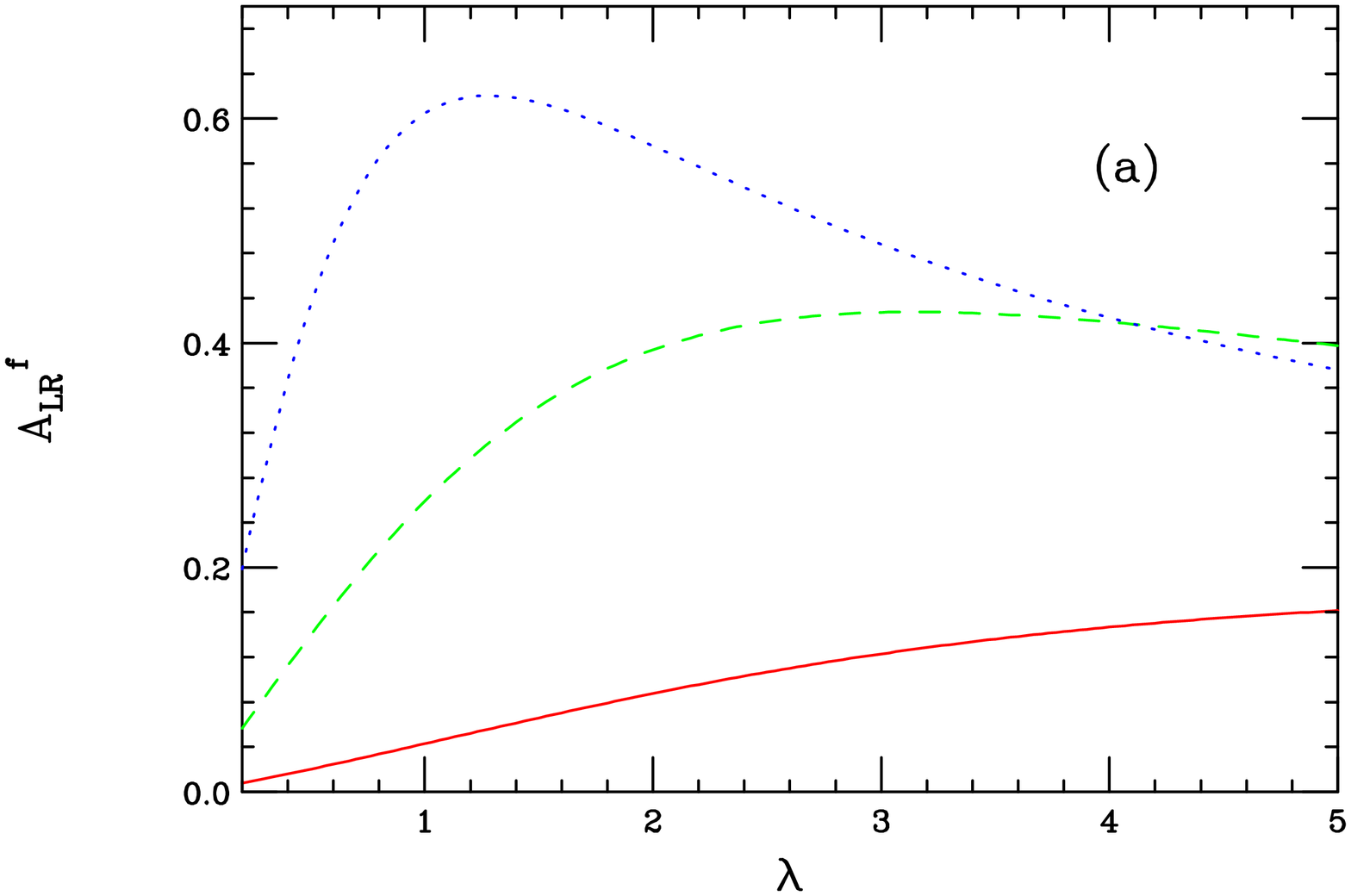,height=8.0cm,width=12cm,angle=90}}
\vspace*{9mm}
\centerline{
\psfig{figure=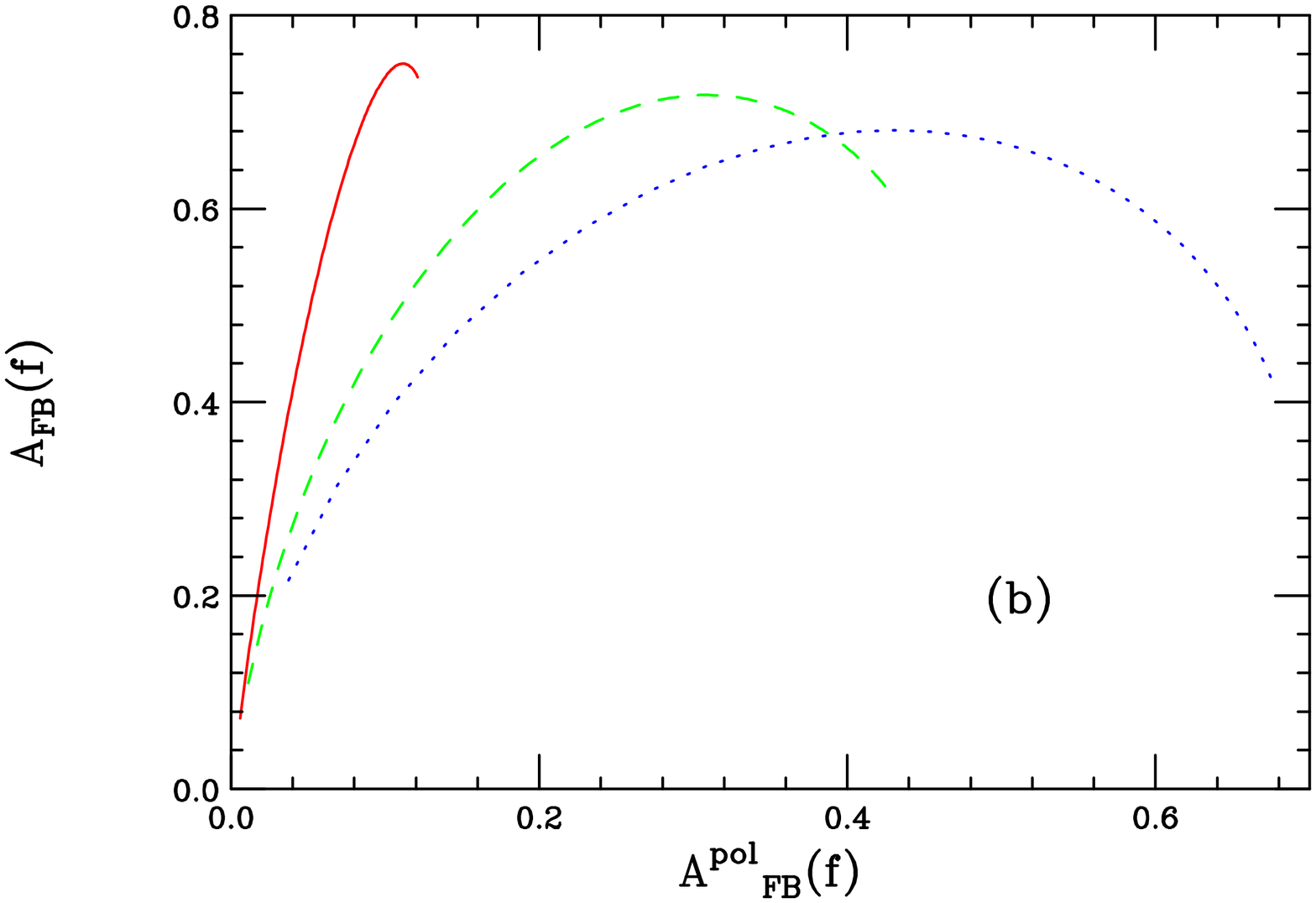,height=8.0cm,width=12cm,angle=90}}
\vspace*{0.0cm}
\caption{(a) $A_{LR}^f$ as a function of the parameter $\lambda$ for 
$f=\ell$(solid), $f=c$(dashed) and $f=b$(dots). (b) Correlations between 
on-peak observables for the same three cases as shown in (a). $\lambda$ 
varies from 0.2 to 5 along each curve.}
\end{figure}
\vspace*{0.4mm}

Of course we need to verify that these single resonance 
relations are numerically badly broken 
before clear experimental signals for more than one resonance can be claimed. 
Statistics will not be a problem with any reasonable integrated luminosity 
since we are sitting on a resonance peak and certainly millions of events 
will be collected. With such large statistics only a small amount of beam 
polarization will be needed to obtain useful asymmetries. 
In principle, to be as model independent as possible in a numerical analysis, 
we should allow the widths $\Gamma_i$ to be greater than or equal to their SM 
values as such heavy KK states may decay to SM SUSY partners as well as to 
presently unknown exotic states. Since the expressions above only depend upon 
the ratio of widths, we let $R=\lambda R_0$ where $R_0$ is the value 
obtained assuming that the KK states have only SM decay 
modes. We then treat $\lambda$ as a free parameter in what follows 
and explore the range 
$1/5 \leq \lambda \leq 5$. Note that as we take $\lambda \to 0(\infty)$ we 
recover the limit corresponding to just a $\gamma^{(1)}(Z^{(1)})$ being 
present.

%
\nn
\begin{figure}[htbp]
\centerline{
\psfig{figure=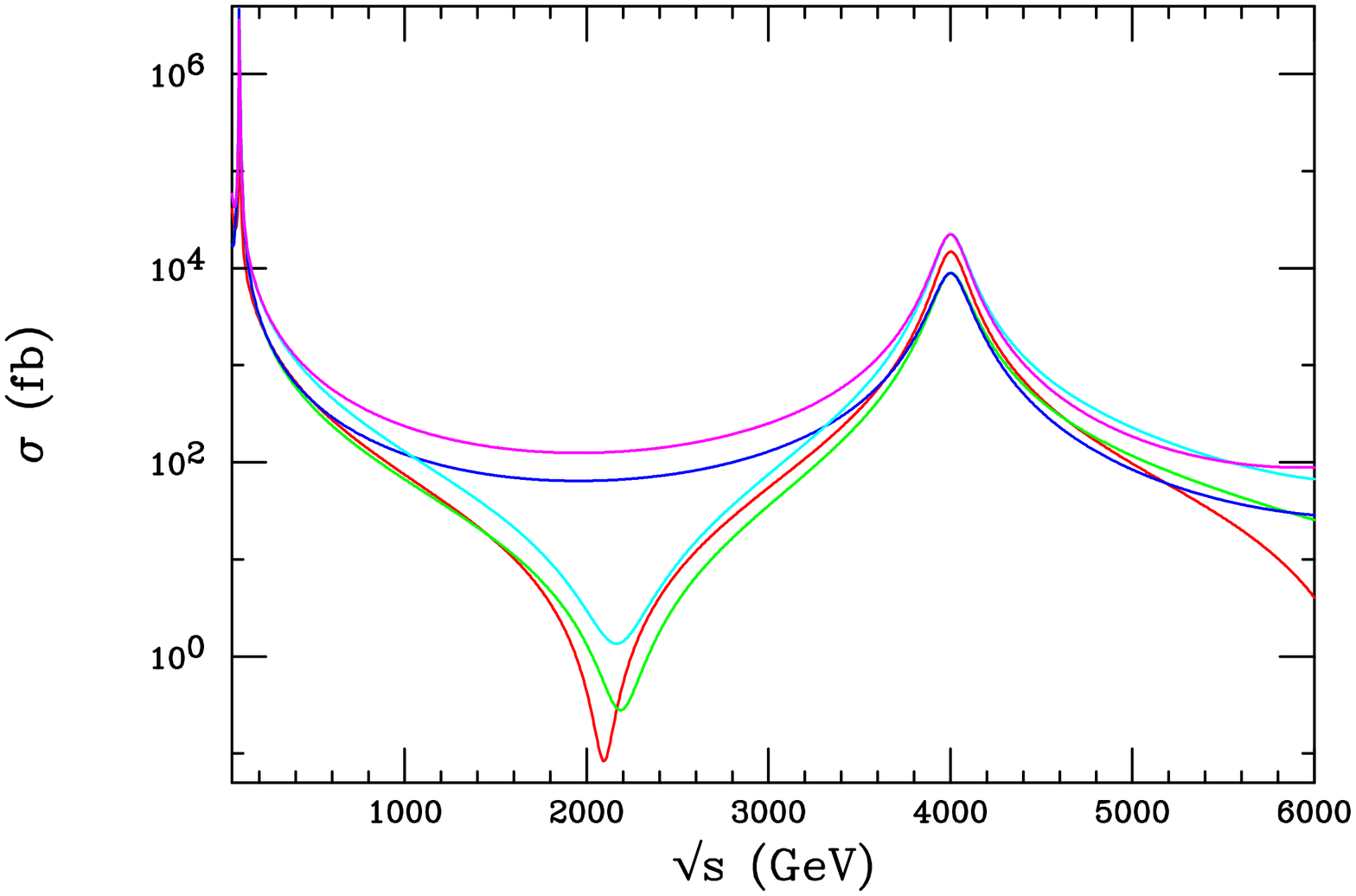,height=8.0cm,width=12cm,angle=90}}
\vspace*{9mm}
\centerline{
\psfig{figure=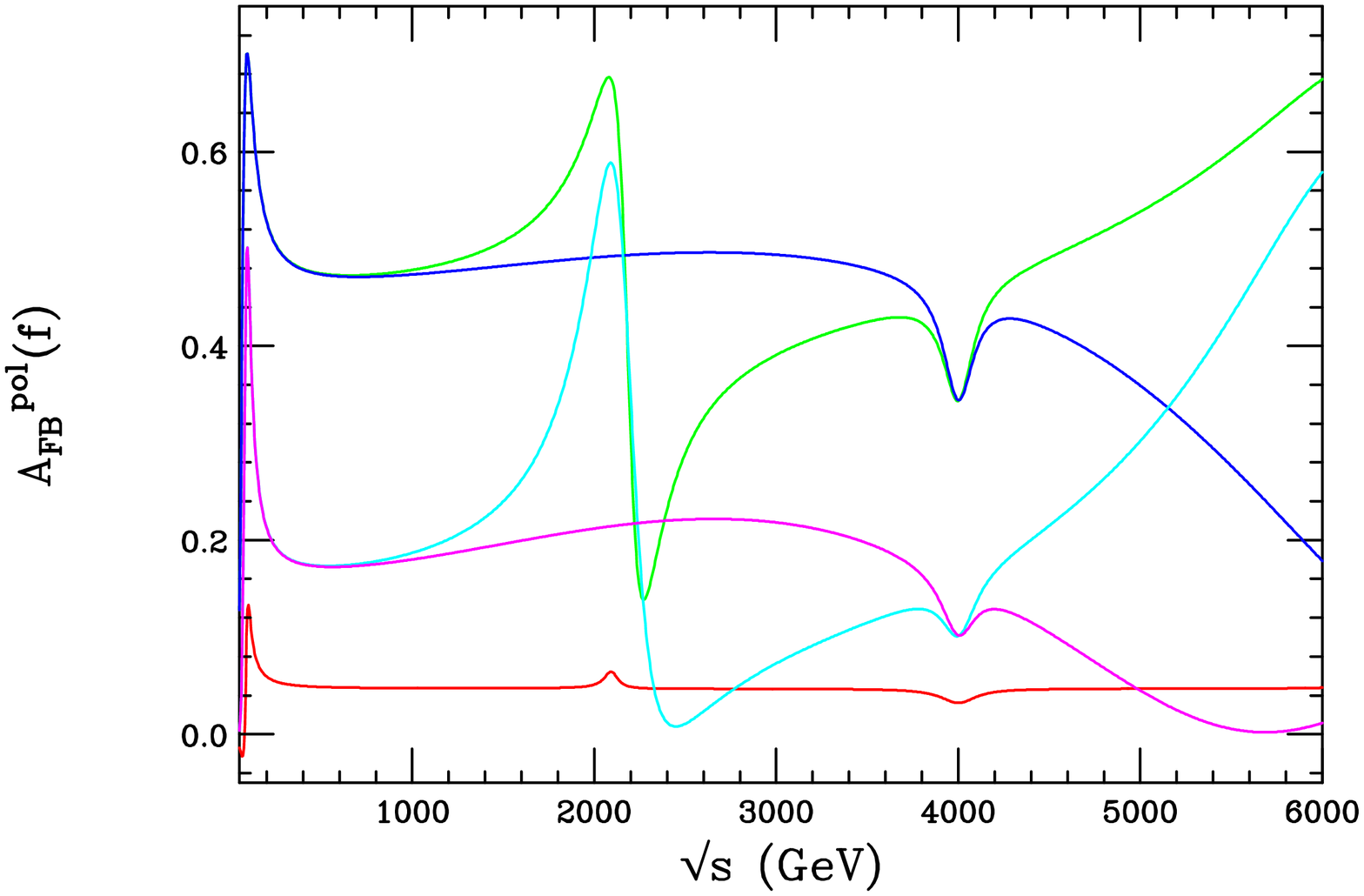,height=8.0cm,width=12cm,angle=90}}
\vspace*{0.0cm}
\caption[*]{Cross sections and polarized $A_{FB}$ for $\mu^+\mu^-\to e^+e^-$
$b\bar b$ and $c\bar c$ as functions of energy in both the 
`conventional' scenario and that of Arkani-Hamed and 
Schmaltz(AS){\cite {schm}} where the quarks and leptons are separated in 
the extra dimension by a distance $D=\pi R$. The red curve 
applies for the $\mu$ final state in either model whereas the green(blue) and 
cyan(magenta) curves label the $b$ and $c$ final states for the 
`conventional'(AS) scenario.}
\end{figure}
\vspace*{0.4mm}

In Fig.4 we display the flavor dependence of $A_{LR}^f$ as a functions of 
$\lambda$. Note that as $\lambda \to 0$ the asymmetries vanish since the 
$\gamma^{(1)}$ has only vector-like couplings. In the opposite limit, for 
extremely large $\lambda$, the $Z^{(1)}$ couplings dominate and a common 
value of $A_{LR}$ will be obtained. It is quite clear, however, 
that over the range of reasonable values of $\lambda$, $A_{LR}^f$ is quite 
obviously flavor dependent. We also show in Fig.4 the correlations between the 
observables $A_{FB}^{pol}(f)$ and $A_{FB}(f)$ which would be flavor 
independent if only a single resonance were present. From the figure we see 
that this is clearly not the case. Note that although $\lambda$ is an 
{\it a priori} unknown parameter, once any one of the electroweak observables 
are measured the value of $\lambda$ will be directly determined. Once $\lambda$ 
is fixed, then the values of all of the other asymmetries, as well as the 
ratios of various partial decay widths, are all completely fixed for the KK 
resonance with uniquely predicted values. This means that we can directly test 
the couplings of this apparent single resonance against what might be 
expected for a degenerate pair of KK excitations without any ambiguities. 

%
\nn
\begin{figure}[htbp]
\centerline{
\psfig{figure=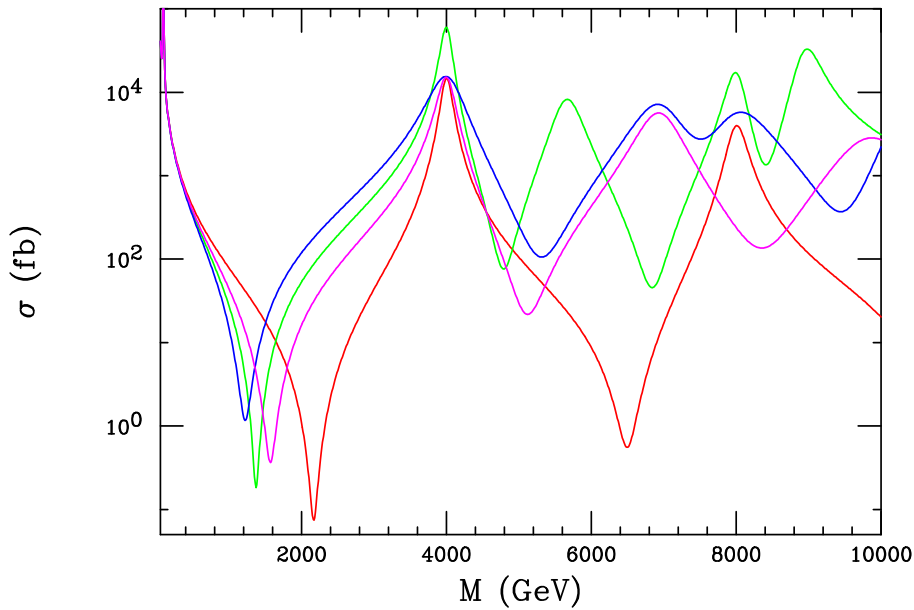,height=8.0cm,width=12cm,angle=90}}
\vspace*{0.0cm}
\caption{Same as Fig. 5a for the process $\mu^+\mu^-\to e^+e^-$ but now also 
including the models listed in Table 1 with $d=2$ assuming $M_1=4$ TeV. 
The red(green,blue,purple) curve corresponds to the $S^1/Z_2$($Z_2\times Z_2$, 
$Z_{3,6}$, $S^2$) compactifications.}
\end{figure}
\vspace*{0.4mm}

In Figs. 5a and 5b we show that although on-resonance measurements of the 
electroweak 
observables, being quadratic in the $Z^{(1)}$ and $\gamma^{(1)}$ couplings, 
will not distinguish between the usual KK scenario and that of the 
Arkani-Hamed and Schmaltz(AS) (whose KK couplings to quarks are of opposite 
sign from the conventional assignments for odd KK levels since quarks and 
leptons are assumed to be separated by a distance $D=\pi R$ in their 
scenario) the data below the peak in the hadronic channel 
will easily allow such a separation. The cross section and asymmetries for 
$\mu^+\mu^-\to e^+e^-$ (or vice versa) is, of course, the same in both cases. 
Such data can be collected by using radiative returns if sufficient luminosity 
is available. The combination 
of on and near resonance measurements will thus completely determine the 
nature of the resonance.

We note that all of the above analysis will go through essentially unchanged 
in any qualitative way 
when we consider the case of the first KK excitation in a theory with more 
than one extra dimension as is shown in Fig.6. Here we see that the shape of 
the excitation curves for the $d=1$ case and the $d>1$ models listed in 
Table 1 will clearly allow the number of dimensions and the compactification 
scheme to be uniquely identified.

\section*{Randall-Sundrum Gravitons at Muon Colliders}

The possibility of extra space-like dimensions with accessible physics near 
the TeV scale has recently opened a new window on the possible 
solutions to the hierarchy problem. Models designed to address this problem 
make use of our ignorance about gravity, in particular,  the fact that gravity 
has yet to be probed at energy scales much above $10^{-3}$ eV in laboratory 
experiments. The prototype scenario in this class of theories is due to 
Arkani-Hamed, Dimopoulos and Dvali(ADD){\cite {nima}} who use the volume 
associated 
with large extra dimensions to bring the $d$-dimensional Planck scale down to 
a few TeV. Here the hierarchy problem is recast into trying to understand the 
rather large ratio of the TeV Planck scale to the size of the extra 
dimensions which may be as large as a fraction of a millimeter. 
The phenomenological{\cite {pheno}} 
implications of this model have been worked out by a large number of authors.
An extrapolation of these analyses to the case of high energy muon colliders 
shows an enormous reach for this kind of physics.

More recently, Randall and Sundrum(RS){\cite {rs}} have proposed a new 
scenario wherein the hierarchy is generated by an
exponential function of the compactification radius, called a warp factor.
Unlike the ADD model, they assume a 5-dimensional non-factorizable geometry, 
based on a slice
of $AdS_5$ spacetime.  Two 3-branes, one being `visible' with the other being
`hidden', with opposite tensions rigidly reside at
$S_1/Z_2$ orbifold fixed points, taken to be $\phi=0,\pi$, where $\phi$ is
the angular coordinate parameterizing the extra dimension.  It is assumed that 
the extra-dimension bulk is only populated by gravity and that the SM lies on 
the brane with negative tension. The 
solution to Einstein's equations for this configuration, maintaining
4-dimensional Poincare invariance, is given by the 5-dimensional metric
\begin{equation}
ds^2=e^{-2\sigma(\phi)}\eta_{\mu\nu}dx^\mu dx^\nu+r_c^2d\phi^2 \,,
\end{equation}
where the Greek indices run over ordinary 4-dimensional spacetime, 
$\sigma(\phi)=kr_c|\phi|$ with $r_c$ being the compactification radius of the
extra dimension, and $0\leq |\phi|\leq\pi$.  Here $k$ is a scale of
order the Planck mass and relates the 5-dimensional Planck scale $M$ to the 
cosmological constant. Examination of the action in the 
4-dimensional effective theory in the RS scenario yields the relationship 
$\mpl^2= M^3/k$ for the reduced effective 4-D Planck scale.  

Assuming that we live on the 3-brane located at $|\phi|=\pi$, it is found
that a field on this brane with the fundamental mass
parameter $m_0$ will appear to have the physical mass $m=e^{-kr_c\pi}m_0$.
TeV scales are thus generated from fundamental scales of order $\mpl$
via a geometrical exponential factor and the observed scale hierarchy is
reproduced if $kr_c\simeq 11-12$.  Hence, due to the exponential nature of the
warp factor, no additional large hierarchies are generated. 

A recent analysis{\cite {dhr}} examined the phenomenological 
implications and constraints on the RS model that arise from the exchange of 
weak scale towers of gravitons. There it was shown that the masses of the KK 
graviton states are given by $m_n=kx_ne^{-kr_c\pi}$ where $x_n$ are the roots 
of $J_1(x_n)=0$, the ordinary Bessel function of order 1. It is important to 
note that these roots are {\it not} 
equally spaced, in contrast to most KK models with one extra dimension, due to 
the non-factorizable metric. 
Expanding the graviton field into the KK states one finds the interaction 
\begin{equation}
{\cal L} = - {1\over\mpl}T^{\alpha\beta}(x)h^{(0)}_{\alpha\beta}(x)-
{1\over\Lambda_\pi}T^{\alpha\beta}(x)\sum_{n=1}^\infty 
h^{(n)}_{\alpha\beta}(x)\,.
\label{effL}
\end{equation}
Here, $T^{\alpha \beta}$ is the stress energy tensor on the brane and we see 
that the zero mode separates from the sum and couples with the usual
4-dimensional strength, $\mpl^{- 1}$; however, all the 
massive KK states are only suppressed by $\Lambda_\pi^{- 1}$, where we find
that $\Lambda_\pi = e^{- kr_c\pi} \mpl$, which is of order  the weak 
scale. 

%
\nn
\begin{figure}[htbp]
\centerline{
\psfig{figure=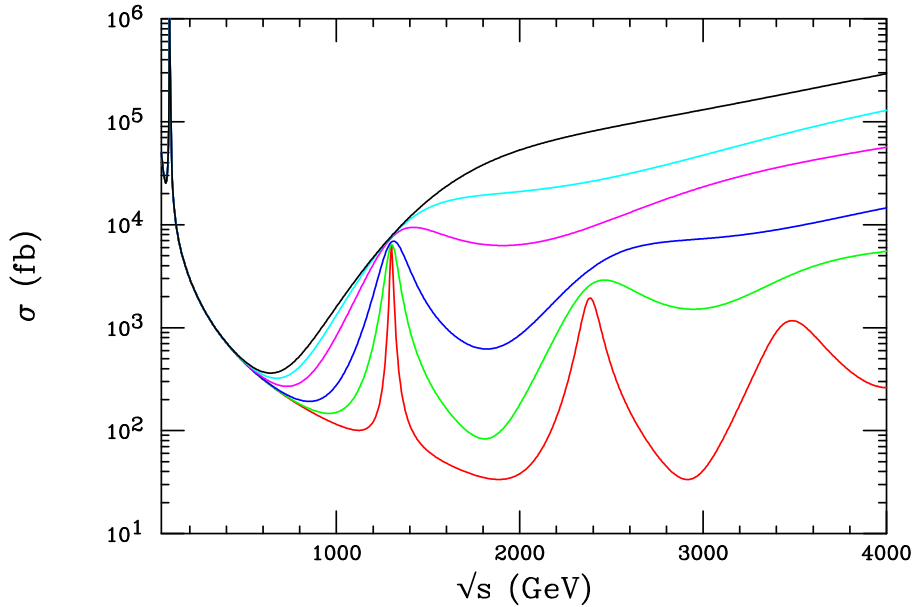,height=8.0cm,width=12cm,angle=90}}
\vspace*{0.0cm}
\caption{Cross section for $\mu^+\mu^-\to e^+e^-$ including the exchange of 
KK gravitons, taking the mass of the first mode to be 1.2 TeV, as a function 
of energy. From top to bottom the curves correspond to c=1.0, 0.7, 0.5, 
0.3, 0.2, and 0.1.}
\end{figure}
\vspace*{0.4mm}

This model has essentially 2 free parameters which we can take to be the mass 
of the first KK graviton mode and the ratio $c=k/\mpl$; the later quantity is 
restricted to be less than unity to maintain the self-consistency of the 
scenario (to prevent a radius of curvature smaller than the Planck scale in 
5 dimensions) and if it is taken too small another hierarchy is formed. 
Figs.7 and 8 show the cross section and $A_{FB}$ for the process 
$\mu^+\mu^- \to e^+e^-$ as a function of $\sqrt s$ in the presence of KK 
graviton resonances for several values of the parameter $c$. For large $c$ 
one does not see the individual resonance structures (since the theory is 
strongly coupled and they are smeared 
together by their large widths which grow as $\sim c^2$) but only a very large 
shoulder somewhat similar to a contact interaction. For small $c$ one sees 
the individual resonances with their widths growing rapidly with 
increasing mass as $\sim m_n^3$. Note that for large $\sqrt s$ where graviton 
exchange dominates the value of $A_{FB}$ is driven to zero. Sitting on any of 
these KK resonances, in the case of small values of $c$, will immediately 
reveal the unique quartic angular distribution corresponding to spin-2 
graviton exchange for the fermions in the 
final state $\sim 1-3\cos^2 \theta +4\cos^4 \theta$.

%
\nn
\begin{figure}[htbp]
\centerline{
\psfig{figure=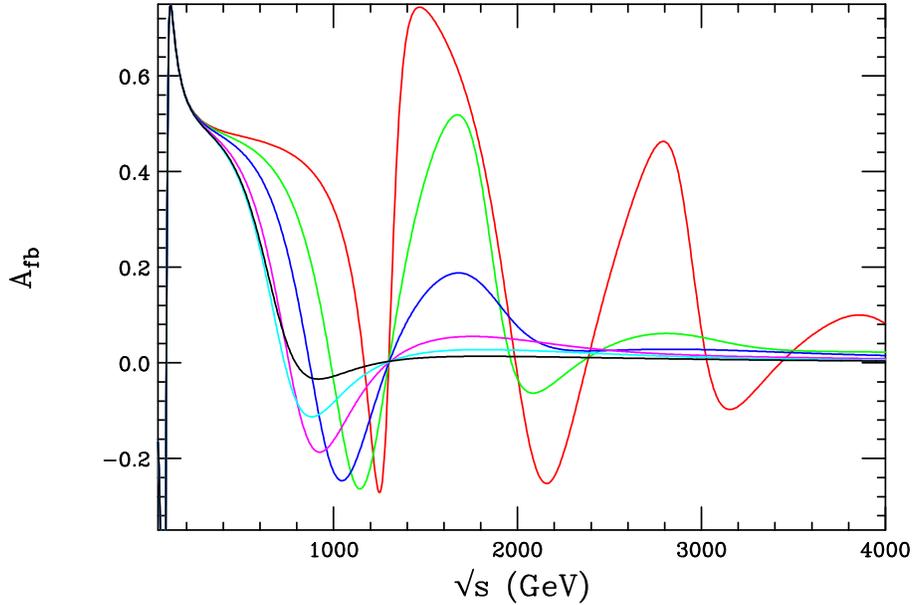,height=8.0cm,width=12cm,angle=90}}
\vspace*{0.0cm}
\caption{Same as the previous figure but now for the Forward-Backward 
asymmetry in the RS model. The color code is as in the previous figure.}
\end{figure}
\vspace*{0.4mm}

\section*{Conclusions}

Present data indicates that the masses of KK excitations of the SM gauge 
bosons must be rather heavy, \eg, $>3.9$ TeV if $d=1$. We have found that:

\begin{itemize}

\item  With an integrated luminosity of $100~fb^{-1}$, the LHC will be able 
to observe KK excitations in the mass range below 
$\simeq 6$ TeV but may not see any KK excitations when $d>1$ since they are 
likely to be more massive. The LHC will not see the second set of 
KK resonances even when $d=1$.

\item  The LHC cannot separate the KK states $\gamma^{(1)}$ from $Z^{(1)}$ 
which will appear together as a single resonance, nor can it obtain 
significant coupling constant information.

\item  The LHC cannot see the $g^{(1)}$ if its mass is in excess of $\sim 4$ 
TeV due to its large width and the energy resolution of the LHC detectors.

\item  The LHC cannot distinguish an extended electroweak model with a 
degenerate $Z'/W'$ from a KK scenario. All we will know is the mass of these 
resonances. 

\item  A LC with $\sqrt s=0.5-1$ TeV will be sensitive to the existence of KK 
states with masses more than an order of magnitude larger than $\sqrt s$ for 
reasonable integrated luminosities $\simeq 100~fb^{-1}$.

\item  At a LC, the extraction of the couplings of an apparent $Z'$, whose 
mass is known from measurements obtained at the LHC, can be performed in a 
straightforward manner with reasonable integrated luminosities. However, the 
$Z'$ hypothesis will yield a poor fit to 
the data if the state in question is actually the 
combined $\gamma^{(1)}/Z^{(1)}$ KK excitation. The LC will not be able to 
identify this state as such--only prove it is not a $Z'$. 

\item  A Muon Collider operating at or above the first KK resonance pole will 
identify it as a KK state provided polarized beams are available.

\item  Measurements of the KK excitation spectrum at Muon Colliders will be 
able to tell us both the number of extra dimensions and how they are 
compactified thus possibly revealing the basic underlying theory upon which 
the KK scenario is based. 

\item  KK excitations of gravitons in the RS model can be studied in detail at 
both LC and Muon Colliders with Muon Colliders providing a much larger reach 
in explorable parameter space. These measurements can completely determine all 
of the parameters of this model.

\end{itemize}

Muon Colliders clearly offer a very important window into the physics of 
Kaluza-Klein excitations.

%
\def\MPL #1 #2 #3 {Mod. Phys. Lett. {\bf#1},\ #2 (#3)}
\def\NPB #1 #2 #3 {Nucl. Phys. {\bf#1},\ #2 (#3)}
\def\PLB #1 #2 #3 {Phys. Lett. {\bf#1},\ #2 (#3)}
\def\PR #1 #2 #3 {Phys. Rep. {\bf#1},\ #2 (#3)}
\def\PRD #1 #2 #3 {Phys. Rev. {\bf#1},\ #2 (#3)}
\def\PRL #1 #2 #3 {Phys. Rev. Lett. {\bf#1},\ #2 (#3)}
\def\RMP #1 #2 #3 {Rev. Mod. Phys. {\bf#1},\ #2 (#3)}
\def\NIM #1 #2 #3 {Nuc. Inst. Meth. {\bf#1},\ #2 (#3)}
\def\ZPC #1 #2 #3 {Z. Phys. {\bf#1},\ #2 (#3)}
\def\EJPC #1 #2 #3 {E. Phys. J. {\bf#1},\ #2 (#3)}
\def\IJMP #1 #2 #3 {Int. J. Mod. Phys. {\bf#1},\ #2 (#3)}

\end{document}